\documentclass[conference]{IEEEtran}
\IEEEoverridecommandlockouts
\usepackage{cite}
\usepackage{amsmath,amssymb,amsfonts}
\usepackage{algorithmic}
\usepackage{graphicx}
\usepackage{textcomp}
\usepackage{xcolor}
\usepackage{subfig}
\usepackage{multirow}
\usepackage{stfloats}
\usepackage{siunitx}
\usepackage{float}
\usepackage{bm}
\usepackage{svg} 
\def\BibTeX{{\rm B\kern-.05em{\sc i\kern-.025em b}\kern-.08em
    T\kern-.1667em\lower.7ex\hbox{E}\kern-.125emX}}
\usepackage[letterpaper, top=0.75in, bottom=1.1in, left=0.625in, right=0.625in]{geometry}
\begin{document}

\title{Fractional Delay Alignment Modulation for Spatially Sparse Wireless Communications\thanks{This work was supported by the National Natural Science Foundation of China under Grant 62071114,  by the "Program for Innovative Talents and Entrepreneur in Jiangsu" under grant number 1104000402, and also by the Outstanding Projects of Overseas Returned Scholars of Nanjing under grant 1104000396.}}

\author{\IEEEauthorblockN{Zhiwen Zhou\IEEEauthorrefmark{1}, Zhiqiang Xiao\IEEEauthorrefmark{1}\IEEEauthorrefmark{2}, and Yong Zeng\IEEEauthorrefmark{1}\IEEEauthorrefmark{2}}
	
\IEEEauthorblockA{\IEEEauthorrefmark{1}National Mobile Communications Research Laboratory, Southeast University, Nanjing 210096, China}
\IEEEauthorblockA{\IEEEauthorrefmark{2}Purple Mountain Laboratories, Nanjing 211111, China}

	\IEEEauthorblockA{Email: \{220220790, zhiqiang\underline{~}xiao, yong\underline{~}zeng\}@seu.edu.cn.}
}

\maketitle

\setlength{\skip\footins}{2ex}

\begin{abstract}
Delay alignment modulation (DAM) is a novel transmission technique for wireless systems with high spatial resolution by leveraging \emph{delay compensation} and \emph{path-based beamforming}, to mitigate the inter-symbol interference (ISI) without resorting to complex channel equalization or multi-carrier transmission. However, most existing studies on DAM consider a simplified scenario by assuming that the channel multi-path delays are integer multiples of the signal sampling interval. This paper investigates DAM for the more general and practical scenarios with fractional multi-path delays. We first analyze the impact of fractional multi-path delays on the existing DAM design, termed \emph{integer DAM} (iDAM), which can only achieve delay compensations that are integer multiples of the sampling interval. It is revealed that the existence of fractional multi-path delays renders iDAM no longer possible to achieve perfect delay alignment. To address this issue, we propose a more generic DAM design called \emph{fractional DAM} (fDAM), which achieves fractional delay pre-compensation via upsampling and fractional delay filtering. By leveraging the \emph{Farrow} filter structure, the proposed approach can eliminate ISI without real-time computation of filter coefficients, as typically required in traditional channel equalization techniques. Simulation results demonstrate that the proposed fDAM outperforms the existing iDAM and orthogonal frequency division multiplexing (OFDM) in terms of symbol error rate (SER) and spectral efficiency, while maintaining a comparable peak-to-average power ratio (PAPR) as iDAM, which is considerably lower than OFDM.
\end{abstract}

\begin{IEEEkeywords}
Delay alignment modulation; fractional delay; single-carrier communication; OFDM
\end{IEEEkeywords}

\section{Introduction}

Inter-symbol interference (ISI) is a critical issue for broadband communications systems under time-dispersive multi-path channels. To resolve this issue, numerous ISI mitigation techniques have been proposed, such as channel equalization, RAKE receivers, and orthogonal frequency-division multiplexing (OFDM). With the ever-increasing scale of antennas \cite{lu2021communicating} and carrier frequency\cite{rappaport2013millimeter}\cite{akyildiz2014terahertz}, wireless channels are becoming increasingly sparse \cite{rangan2014millimeter}\cite{THz_ch2}. Recently, by exploiting the channel spatial sparsity and high spatial resolution of large antenna arrays, a novel transmission technique termed delay alignment modulation (DAM) was proposed\cite{lu2022delay}\cite{lu2023delay}. Different from conventional methods such as channel equalization or OFDM, DAM leverages two key techniques to eliminate the ISI, i.e., delay compensation and path-based beamforming. Thus, DAM has the advantages of smaller guard interval overhead, lower peak-to-average-power ratio (PAPR), and higher Doppler resilience compared to OFDM \cite{zhang2023delay}.

However, most existing studies on DAM consider a simplified scenario where the multi-path channel delays are assumed to be integer multiples of the signal sampling interval \cite{lu2022delay}\cite{lu2023delay}. In this case, to achieve perfect DAM so that all multi-path signals arrive at the receiver simultaneously and constructively, the transmitter only needs to deliberately introduce delay compensations that are also integer multiples of the sampling interval. We term such existing DAM designs \cite{lu2022delay}\cite{lu2023delay} as \emph{integer DAM} (iDAM).  However, for practical scenarios, the multi-path delays are not necessarily integer multiples of the sampling interval. Thus, if the existing iDAM design is directly applied, there may exist residual fractional delays, whose length is non-negligible as compared to the symbol duration in single-carrier systems. Specifically, the resulting residual fractional delays after the conventional iDAM can lead to time misalignment, causing residual ISI \cite{frac_sc}. To tackle this issue, a tap-based DAM design was proposed in \cite{ding2022channel}, where delay compensation and beamforming are performed based on the delay taps, with each tap containing multiple physical paths. However, fractional multi-path delays can still cause the delay tap power leakage issue \cite{xiao2023integrated}, which may result in strong spatial correlation among adjacent delay taps, rendering the tap-based zero-forcing (ZF) and maximum ratio transmission (MRT) beamforming ineffective.
\vspace*{-1pt}

In this paper, we investigate DAM for a more general and practical wireless communication system with fractional multi-path delays. First, we study the impact of fractional multi-path delays on the existing iDAM designs\cite{lu2022delay}\cite{lu2023delay}. The analytic results demonstrate that in the presence of fractional delays, although iDAM can still significantly reduce the channel delay spread, perfect delay alignment is no longer possible. Thus, the resulting residual ISI after iDAM may degrade the system performance. To address this issue, we propose a more generic DAM design, termed \emph{fractional DAM} (fDAM), by noting that perfect delay alignment is still doable even with fractional multi-path delays, as long as fractional delay compensations can be achieved.  To this end, we employ the upsampling and fractional delay filtering techniques at the transmitter side. By leveraging the \emph{Farrow} filter structure, the proposed approach can achieve fractional delay compensation without real-time computation of filter coefficients, as typically required in traditional channel equalization techniques. Moreover, as compared to the use of high upsampling ratios or dedicated hardware such as true time delay (TTD) lines, the proposed approach achieves fractional delay compensation with a relatively low hardware cost, as it only requires an upsampling ratio of 2. Extensive simulation results demonstrate that the proposed fDAM outperforms the existing iDAM and OFDM in terms of symbol error rate (SER) and spectrum efficiency for scenarios with fractional multi-path delays. Furthermore, the proposed fDAM can achieve a comparable peak-to-average power ratio (PAPR) as iDAM, which is far lower than that of OFDM.
 \vspace*{-2pt}
\section{System Model}\label{sys model}
\begin{figure}[!htbp]
 \vspace*{-13pt}
\centering
\setlength{\abovecaptionskip}{0pt}
\setlength{\belowcaptionskip}{0pt}
    \includegraphics[width=0.38\textwidth]{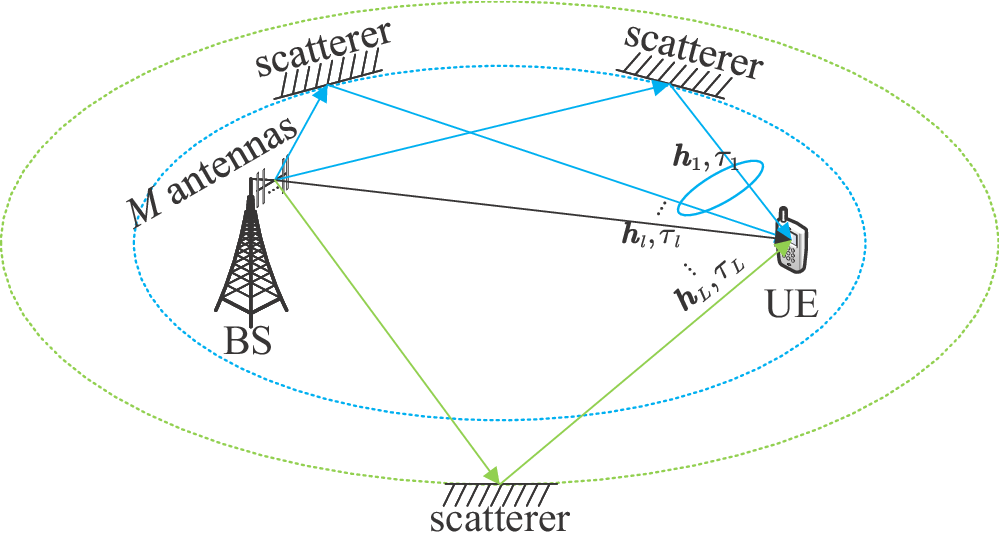}
	\hfil
\caption{A spatially sparse wireless communication system with fractional multi-path delays.}
	\label{scene}
\vspace*{-12pt}
\end{figure}

As shown in Fig. \ref{scene}, a spatially sparse multiple-input single-output (MISO) wireless communication system is considered, where a base station (BS) equipped with $M\gg1$ transmit antennas wishes to serve a single-antenna user. Let $L$ denote the number of physical channel paths, which is much smaller than $M$ for spatially sparse systems. Thus, the channel impulse response is written as
\begin{equation}
\abovedisplayshortskip=2pt
\belowdisplayshortskip=2pt
\abovedisplayskip=2pt
\belowdisplayskip=2pt
\begin{aligned}
   \bm{h}^H\left(t\right)=\sum_{l=1}^{L}{\bm{h}_l^H\delta\left(t-\tau_l\right)},
\end{aligned}
\label{h_t}
\end{equation}
where $\bm{h}_l \in \mathbb{C} ^{M\times 1}$ and $\tau_l \ge 0$ denote the channel coefficient vector and the propagation delay of the $l$th path, respectively.

Note that different from most existing works on DAM \cite{lu2022delay,lu2023delay,zhang2023delay} which assume that the multi-path delays are integer multiples of $T$, where $T$ is the symbol duration, in this paper, we consider a more general and practical case without making the above assumption. In this case, the delay of the $l$th path $\tau_l$ can be expressed as
\begin{equation}
\abovedisplayshortskip=2pt
\belowdisplayshortskip=2pt
\abovedisplayskip=2pt
\belowdisplayskip=2pt
\begin{aligned}
\tau _l=n_lT+\tau _{f,l},
\end{aligned}
\label{delay_decompose}
\end{equation}
where $n_l=\left. \lfloor \left. \frac{\tau _l}{T} \right. \rceil\right.$ and $\tau_{f,l}=\left(\tau_l-n_lT\right)\in\left[\left.-\frac{T}{2},\frac{T}{2}\right)\right.$ are the integer and fractional parts of the delay, respectively, with $\lfloor\cdot\rceil$ denoting round to the nearest integer.
\begin{figure}[!htbp]
\centering
\setlength{\abovecaptionskip}{0pt}
\setlength{\belowcaptionskip}{0pt}
    \includegraphics[width=0.38\textwidth]{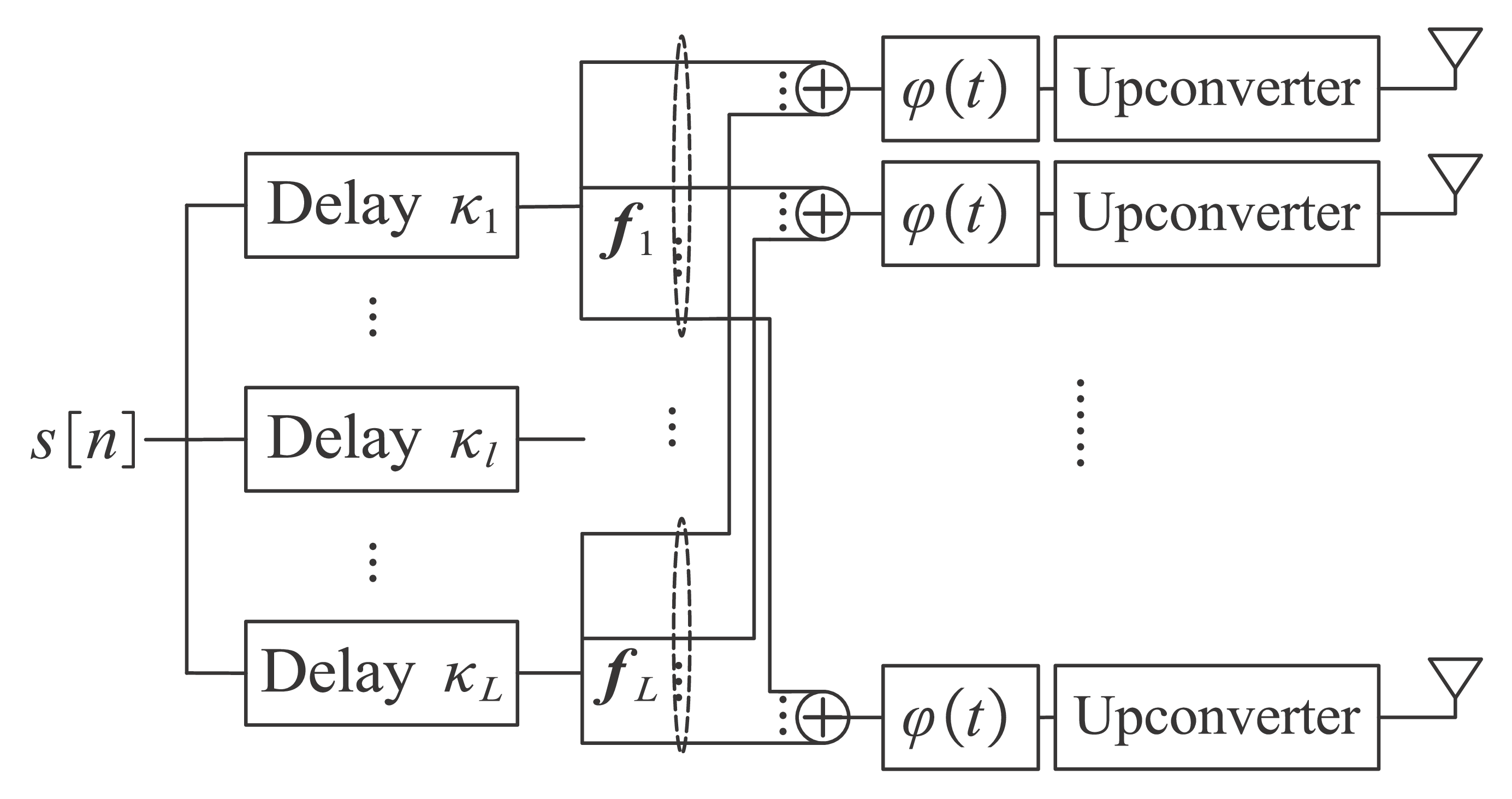}
	\hfil
\caption{Transmitter structure of existing iDAM \cite{lu2022delay}.}
	\label{fig_iDAM_structure}
 \vspace*{-23pt}
\end{figure}
As shown in Fig. \ref{fig_iDAM_structure}, with the existing iDAM design proposed in \cite{lu2022delay}, the transmitted baseband discrete-time signal is
\begin{equation}
\abovedisplayshortskip=4pt
\belowdisplayshortskip=4pt
\abovedisplayskip=4pt
\belowdisplayskip=4pt
\begin{aligned}
    \bm{x}\left[n\right]=\sum_{l=1}^{L}{\bm{f}_ls\left[n-\kappa_l\right]},
\end{aligned}
\label{tx_baseband_dig}
\end{equation} 
where $ \{{s\left[n\right]}\}_{n=-\infty}^\infty\ $ are the independent and identically distributed (i.i.d) information-bearing symbols with normalized power $\mathbb{E} \{ \left| s\left[ n \right] \right|^2 \} =1$, $ \bm{f}_l \in \mathbb{C} ^{M\times 1} $ and $ \kappa_l \in \mathbb{C} ^{M\times 1} $ are the path-based beamforming vector and the deliberately introduced delay compensation for the $l$th path, respectively. Note that $ \sum_{l=1}^L{\left\| \boldsymbol{f}_l \right\| ^2}\le PT $, where $P$ is the maximum transmit power. For the conventional iDAM, $ \{\kappa_l\}_{l=1}^L\ $ are non-negative integers. Let $ \varphi\left(t\right) $ denote the transmit pulse shaping filter. Then the transmitted baseband continuous-time signal is 
\begin{equation}
\abovedisplayshortskip=2pt
\belowdisplayshortskip=2pt
\abovedisplayskip=2pt
\belowdisplayskip=2pt
\begin{aligned}
    \bm{x}\left(t\right)=\sum_{n=-\infty}^{\infty}\bm{x}\left[n\right]\varphi\left(t-nT\right).
\end{aligned}
\label{tx_baseband_sig}
\end{equation}
By substituting (\ref{tx_baseband_dig}) into (\ref{tx_baseband_sig}), we have
\begin{equation}
\abovedisplayshortskip=2pt
\belowdisplayshortskip=2pt
\abovedisplayskip=2pt
\belowdisplayskip=2pt
\begin{aligned}
    \bm{x}\left(t\right)&=\sum_{n=-\infty}^{\infty}{\sum_{l=1}^{L}{\bm{f}_ls\left[n\right]}\varphi\left(t-nT-\kappa_lT\right)}
    \\
    &=\sum_{l=1}^{L}\bm{f}_ls\left(t-\kappa_lT\right),
\end{aligned}
\label{tx_baseband_sig_2}
\end{equation}
where $s\left(t\right)$ is given by
\begin{equation}
\abovedisplayshortskip=2pt
\belowdisplayshortskip=2pt
\abovedisplayskip=2pt
\belowdisplayskip=2pt
\begin{aligned}
    s\left(t\right)\triangleq\sum_{n=-\infty}^{\infty}s\left[n\right]\varphi\left(t-nT\right).
\end{aligned}
\label{s_t}
\end{equation}
Thus, with the channel impulse response in (\ref{h_t}), the signal received at the receiver is
\begin{equation}
\abovedisplayshortskip=2pt
\belowdisplayshortskip=2pt
\abovedisplayskip=2pt
\belowdisplayskip=2pt
\begin{aligned}
    y\left(t\right)&=\bm{h}^H\left(t\right)\ast\bm{x}\left(t\right)+z\left(t\right)
    \\
    &=\sum_{l=1}^{L}{\bm{h}_l^H\bm{f}_ls\left(t-\kappa_lT-\tau_l\right)}
    \\
    &+\sum_{l=1}^{L}\sum_{l^\prime\neq l}^{L}{\bm{h}_l^H\bm{f}_{l^\prime}s\left(t-\kappa_{l^\prime}T-\tau_l\right)}+z\left(t\right),
\end{aligned}
\label{y_t}
\end{equation}
where $z\left( t \right)$ is the AWGN with power spectral density $\sigma ^2$.
\vspace*{-8pt}
\section{Impact of Fractional Delays on iDAM}\label{impact}

For the existing iDAM design\cite{lu2022delay}\cite{lu2023delay}, it is assumed that the multi-path delays are integer multiples of $T$, i.e., $\tau_{f,l}=0, \forall l$ in (\ref{delay_decompose}). In this case, the deliberately introduced delay compensation in (\ref{tx_baseband_dig}) can be set as $\kappa_l = n_{\max} - n_l \ge 0$, with $n_{\max}=\underset{l}{\max}\left\{ n_l \right\}, l = 1,\ldots,L$. In this section, we aim to study the impact of fractional multi-path delays if the existing iDAM design is directly applied. In this case, by substituting $\tau_l$ in (\ref{delay_decompose}) and the integer $\kappa_l$ design into (\ref{y_t}), we have
\begin{equation}
\abovedisplayshortskip=4pt
\belowdisplayshortskip=4pt
\abovedisplayskip=4pt
\belowdisplayskip=4pt
\begin{aligned}
y\left( t \right) &= \sum_{l=1}^L{\bm{h}_{l}^{H}\bm{f}_l}s\left( t-n_{\max}T-\tau _{f,l} \right)
\\
&+\sum_{l=1}^L{\sum_{l^{\prime}\ne l}^L{\bm{h}_{l}^{H}\bm{f}_{l^{\prime}}s\left( t-n_{\max}T+n_{l^{\prime}}T-\tau _l \right)}}+z\left( t \right) .
\end{aligned}
\label{y_t_full}
\end{equation}
With the path-based ZF beamforming for $\{\bm{f}_l\}_{l=1}^{L}$\cite{lu2022delay}, which ensures that  $\bm{h}_l^H\bm{f}_{l^\prime}=0,\forall l\neq l^\prime$, (\ref{y_t_full}) reduces to
\begin{equation}
\abovedisplayshortskip=2pt
\belowdisplayshortskip=2pt
\abovedisplayskip=2pt
\belowdisplayskip=2pt
\begin{aligned}
y\left( t \right) &= \sum_{l=1}^L{\bm{h}_{l}^{H}\bm{f}_l}s\left( t-n_{\max}T-\tau _{f,l} \right) +z\left(t\right)
\\
&= \sum_{l=1}^L{\bm{h}_{l}^{H}\bm{f}_l}\sum_{n=-\infty}^{\infty}{s}\left[ n \right] \varphi \left( t-nT-\tau _{f,l}-n_{\max}T \right)
\\
&+z\left(t\right).
\end{aligned}
\label{y_t_zf}
\end{equation}
By applying the receiving matched filter $\varphi\left(t\right)$ to $y(t)$, we have
\begin{equation}
\abovedisplayshortskip=2pt
\belowdisplayshortskip=2pt
\abovedisplayskip=2pt
\belowdisplayskip=2pt
\begin{aligned}
r\left(t\right)&=y\left(t\right)\ast\varphi\left(t\right)
\\
&=\sum_{l=1}^{L}{\bm{h}_l^H\bm{f}_l}\sum_{n=-\infty}^{\infty}s\left[n\right]\varrho \left(t-nT-\tau _{f,l}-n_{\max}T \right)
\\
&+v\left(t\right),
\end{aligned}
\label{r_t}
\end{equation}
where $\varrho  \left( t \right) \triangleq \varphi \left( t \right) *\varphi \left( t \right)$ denotes the matched filter response, and $v\left( t \right) \triangleq z\left( t \right) *\varphi \left( t \right)$ is the resulting noise after receiving matched filtering (MF).

For the receiving MF output $r\left(t\right)$, by sampling at $t=kT+n_{\max}T+T_\mathrm{peak}$ with integer $k$, we have
\begin{equation}
\abovedisplayshortskip=2pt
\belowdisplayshortskip=2pt
\abovedisplayskip=2pt
\belowdisplayskip=2pt
\begin{aligned}
r\left[k\right]&=\sum_{l=1}^{L}{\bm{h}_l^H\bm{f}_l}\sum_{n=-\infty}^{\infty}s\left[n\right]\psi\left(\left(k-n\right)T-\tau_{f,l}\right)+v\left[k\right]
\\
&=\underbrace{\left(\sum_{l=1}^{L}{\bm{h}_l^H\bm{f}_l}\psi\left(-\tau_{f,l}\right)\right)s\left[k\right]}_\mathrm{desired\,signal}
\\
&+\underbrace{\sum_{n\neq k}{\left(\sum_{l=1}^{L}{\bm{h}_l^H\bm{f}_l\psi\left(\left(k-n\right)T-\tau_{f,l}\right)}\right)s\left[n\right]}}_\mathrm{ISI}+v\left[k\right],
\end{aligned}
\label{r_k}
\end{equation}
where $\left\{ v[k] \right\} _{k=-\infty}^{\infty}$ are i.i.d complex normal variables satisfying $v[k] \!\sim\! \mathcal{C} \mathcal{N} (0,\sigma ^2)$\cite{proakis2008digital} and $\psi\left(t\right)\!\triangleq\!\varrho \left(t+T_\mathrm{peak}\right)$ with $T_\mathrm{peak}\!=\!\underset{t}{\max}\left\{ \varrho  \left( t \right) \right\}$ being the time instant of the peak MF output. In practice, $\varrho \left(t\right)$ must be causal, thus we have $T_\mathrm{peak}\!>\!0$. The pulse $\psi\left(t\right)$ satisfies the Nyquist ISI-free property $\psi\left(nT\right)\!=\!\delta \left[n\right]$, where $\delta[\cdot]$ denotes the Kronecker delta function.

Note that different from \cite{lu2022delay,lu2023delay,zhang2023delay}, due to the existence of fractional delays $\tau_{f,l}$, there exists residual ISI even after iDAM, as shown in the second term of (\ref{r_k}). By noting that $s[n]$ are i.i.d. with normalized power, the output signal-to-interference-plus-noise ratio (SINR) is
\begin{equation}
\abovedisplayshortskip=2pt
\belowdisplayshortskip=2pt
\abovedisplayskip=2pt
\belowdisplayskip=2pt
\begin{aligned}
\gamma 
&=\frac{\left|\sum_{l=1}^{L}{\bm{h}_l^H\bm{f}_l}\psi\left(-\tau_{f,l}\right)\right|^2}{\sum_{n\neq k}\left|\sum_{l=1}^{L}{\bm{h}_l^H\bm{f}_l}\psi\left(\left(k-n\right)T-\tau_{f,l}\right)\right|^2+\sigma^2}.
\end{aligned}
\label{SINR}
\end{equation}

If the multi-path delays were integer multiples of $T$, as assumed in \cite{lu2022delay,lu2023delay,zhang2023delay}, i.e. $\tau _{f,l}=0,\forall l$, then by using the property $\psi\left(nT\right)=\delta \left[n\right]$, (\ref{SINR}) reduces to
\begin{equation}
\abovedisplayshortskip=2pt
\belowdisplayshortskip=2pt
\abovedisplayskip=2pt
\belowdisplayskip=2pt
\begin{aligned}
\gamma =\frac{\left| \sum_{l=1}^L{\bm{h}_{l}^{H}\bm{f}_l} \right|^2}{\sigma ^2},
\end{aligned}
\label{SINR2_ideal}
\end{equation}
which is the ideal signal-to-noise ratio (SNR) for DAM without ISI \cite{lu2022delay}. 

However, in practice, due to the existence of fractional multi-path delays, $ \tau_{f,l}\neq 0 $, the ISI term is non-zero. Moreover, due to the deviated sampling time $-\tau_{f,l}$, the desired signal power in the numerator of (\ref{SINR}) is also compromised. Thus, if the conventional iDAM is directly applied in practical systems with fractional multi-path delays, it will result in performance degradation, which will be verified via simulations in Section \ref{Sim}. To address this issue, in the following, we propose a novel DAM design called fDAM.
\vspace*{-3pt}
\section{Fractional DAM}\label{fDAM}

Intuitively, in order to achieve perfect delay alignment even with fractional multi-path delays, the transmitter needs to realize fractional delay compensations. Thus, the transmitted signal in (\ref{tx_baseband_sig_2}) should be designed as
\begin{equation}
\abovedisplayshortskip=2pt
\belowdisplayshortskip=2pt
\abovedisplayskip=2pt
\belowdisplayskip=2pt
\begin{aligned}
    \bm{x}\left(t\right)&=\sum_{l=1}^{L}\bm{f}_ls\left(t-\rho_l\right),
\end{aligned}
\label{tx_baseband_sig_fraction_delay}
\end{equation}
where $\rho_l=\tau _{\max}-\tau_l$ denotes the fractional delay compensation, with $\tau _{\max}\triangleq\underset{l}{\max}\left\{ \tau _l \right\}, l=1,\ldots,L$. Note that different from (\ref{tx_baseband_sig_2}), $\rho_l$ may not be integer multiples of $T$. The received signal in (\ref{y_t}) is then rewritten as
\begin{equation}
\abovedisplayshortskip=2pt
\belowdisplayshortskip=2pt
\abovedisplayskip=2pt
\belowdisplayskip=2pt
\begin{aligned}
    y\left(t\right)&=\bm{h}^H\left(t\right)\ast\bm{x}\left(t\right)+z\left(t\right)
    \\
    &=\left(\sum_{l=1}^{L}{\bm{h}_l^H\bm{f}_l}\right)s\left(t-\tau _{\max}\right)
    \\
    &+\sum_{l=1}^{L}\sum_{l^\prime\neq l}^{L}{\bm{h}_l^H\bm{f}_{l^\prime}s\left(t-\tau_{\max}+\tau_{l^\prime}-\tau_l\right)}+z\left(t\right).
\end{aligned}
\label{y_t_fraction_delay}
\end{equation}
Thus, with the path-based ZF beamforming such that $\bm{h}_l^H \bm{f}_l'=0,\forall l\neq l'$, the resulting received signal is
\begin{equation}
\abovedisplayshortskip=2pt
\belowdisplayshortskip=2pt
\abovedisplayskip=2pt
\belowdisplayskip=2pt
\begin{aligned}
    y\left(t\right)=\left(\sum_{l=1}^{L}{\bm{h}_l^H\bm{f}_l}\right)s\left(t-\tau _{\max}\right)+z\left(t\right),
\end{aligned}
\label{y_t_isi_free}
\end{equation}
where different from (\ref{y_t_zf}), the ISI is completely eliminated, thanks to the fractional delay pre-compensation in (\ref{tx_baseband_sig_fraction_delay}).  In this case, we would have the same SNR expression as (\ref{SINR2_ideal}) even with fractional multi-path delays.

However, it is impossible to achieve the desired fractional delay compensation with the existing iDAM design \cite{lu2022delay} shown in Fig.\ref{fig_iDAM_structure}, since the delay compensation therein is performed in the digital domain, which must be integer multiples of the symbol duration $T$. To achieve the desired fractional delay compensation, there exist at least three potential solutions.

One possible solution is to increase the sampling rate at the transmitter, say dozens or even hundreds times the symbol rate, so that the sampling interval is only a small fraction of the symbol duration $T$. As a result, the fractional delay shifts can be well approximated by the digital time shifts with negligible errors. However, this method will drastically increase the hardware requirements of the transmitter, because high-speed digital-to-analog converters (DACs) are required. 

Another potential solution is to move the delay compensation operation from the digital domain to the analog domain. To this end, hardware components like TTD lines could be used to achieve analog time shifts, but they will significantly increase the cost since $L$ TTD lines are required for perfect delay alignment.

Fortunately, by exploiting the fact that the transmitted signal is band-limited, in this paper, we propose an efficient method for achieving fractional delay compensation by the use of digital filters known as fractional delay filters \cite{frac_delay}. These filters are able to achieve fractional time shifts in the digital domain with low complexity and hardware requirements and avoid prohibitive upsampling factors required by the first method mentioned above, or additional hardware components like TTD lines.

\begin{figure}[!htbp]
\centering
\vspace{-12pt}
\setlength{\abovecaptionskip}{0pt}
\setlength{\belowcaptionskip}{-10pt}
    \includegraphics[width=0.42\textwidth]{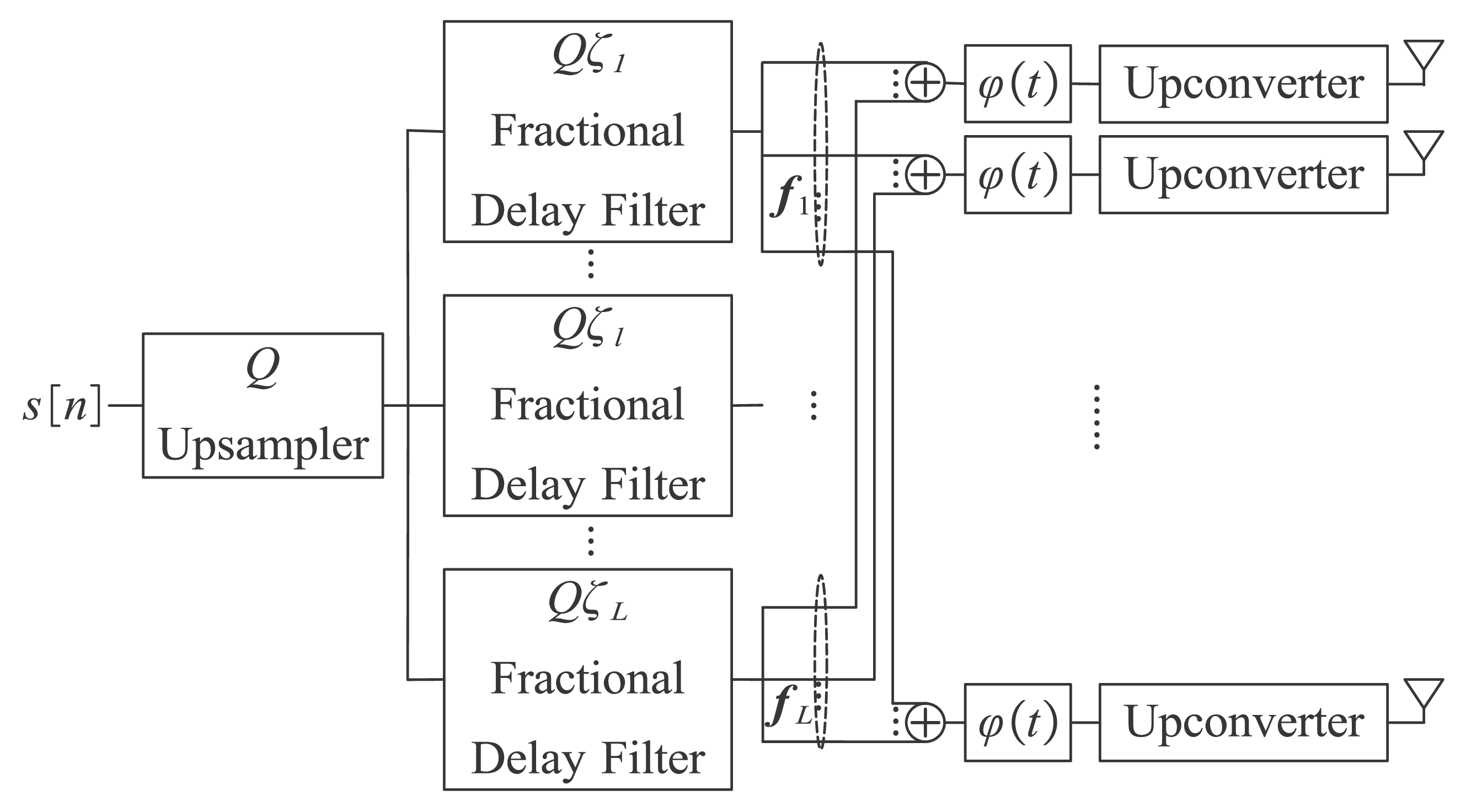}
	\hfil
\caption{Transmitter structure of the proposed fDAM.}
	\label{fig_fDAM_structure}

\end{figure}

Fig. \ref{fig_fDAM_structure} shows the transmitter structure of the proposed fDAM design, where fractional delay compensation is achieved by a combination of upsampling with factor $Q$ and fractional delay filters. In this case, the transmitted discrete-time signal is
\begin{equation}
\abovedisplayshortskip=2pt
\belowdisplayshortskip=2pt
\abovedisplayskip=2pt
\belowdisplayskip=2pt
\begin{aligned}
\bm{x}\left[n\right]=\sum_{l=1}^{L}{\bm{f}_ls_u\left[n\right]}\ast d_{Q\zeta_l}\left[n\right],
\end{aligned}
\label{tx_baseband_dig_frac}
\end{equation}
where $s_u\left[n\right]$ is the upsampled version of $s\left[n\right]$, with an upsampling factor of $Q$, i.e.,
\begin{equation}
\abovedisplayshortskip=2pt
\belowdisplayshortskip=2pt
\abovedisplayskip=2pt
\belowdisplayskip=2pt
\begin{aligned}
s_u\left[ n \right] =\left\{ \begin{aligned}
	&s\left[ n/Q \right] ,&&n\,\,\mathrm{mod} \,\,Q=0\\
	&0,&&\text{otherwise}\\
\end{aligned} \right. ,
\end{aligned}
\label{upsampling}
\end{equation}
and $d_{Q\zeta_l}\left[n\right]$ is the fractional delay filter with desired delay $Q\zeta_lT_s=\zeta_lT$, where $Q\zeta_l$ is in general non-integer and $T_s=T/Q$ is the sampling interval after upsampling. The value for desired $Q$ and $\zeta_l$ will become clear later. {\it Farrow} filter is one of the most popular fractional delay filters, which can achieve delay shifts in real-time \cite{farrow}. Typically, $d_{Q\zeta_l}\left[n\right]$ is designed so that its discrete-time Fourier transform (DTFT) approximates the ideal fractional delay filter with delay $Q\zeta_l$, i.e.,
\begin{equation}
\abovedisplayshortskip=2pt
\belowdisplayshortskip=2pt
\abovedisplayskip=2pt
\belowdisplayskip=2pt
\begin{aligned}
D_{Q\zeta_l}\left(f\right)\approx e^{-j2\pi fQ\zeta_l},\ \ f\in\left[-\frac{1}{2}\right.,\left.\frac{1}{2}\right).
\end{aligned}
\label{frac_filter}
\end{equation}
Note that due to the periodic nature of DTFTs, $D_{Q\zeta_l}\left(f\right)$ can only be designed in one period, thus it is only required for (\ref{frac_filter}) to hold true for $f\in\left[-\frac{1}{2}\right.,\left.\frac{1}{2}\right)$. Taking into account the approximation errors, it is appropriate to assume that for $f\in\left[-0.4\right.,\left.0.4\right)$, the approximation in (\ref{frac_filter}) is accurate \cite{frac_delay}.
With the above designs, the transmitted baseband signal by fDAM can be written as
\begin{equation}
\abovedisplayshortskip=2pt
\belowdisplayshortskip=2pt
\abovedisplayskip=2pt
\belowdisplayskip=2pt
\begin{aligned}
\bm{x}\left(t\right)=\sum_{n=-\infty}^{\infty}\bm{x}\left[n\right]\varphi\left(t-\frac{nT}{Q}\right).
\end{aligned}
\label{tx_baseband_sig_frac}
\end{equation}
For ease of exposition, (\ref{tx_baseband_sig_frac}) can be equivalently written as
\begin{equation}
\abovedisplayshortskip=2pt
\belowdisplayshortskip=2pt
\abovedisplayskip=2pt
\belowdisplayskip=2pt
\begin{aligned}
\bm{x}\left(t\right)=\bm{x}_i\left(t\right)\ast\varphi\left(t\right),
\end{aligned}
\label{tx_baseband_sig_frac2}
\end{equation}
where
\begin{equation}
\abovedisplayshortskip=2pt
\belowdisplayshortskip=2pt
\abovedisplayskip=2pt
\belowdisplayskip=2pt
\begin{aligned}
\bm{x}_i\left(t\right)=\sum_{n=-\infty}^{\infty}\bm{x}\left[n\right]\delta\left(t-\frac{nT}{Q}\right).
\end{aligned}
\label{x_i_t}
\end{equation}
Denote by $f_B=1/T$ the symbol rate. The continuous Fourier transform (CFT) of $\bm{x}_i\left(t\right)$ can be expressed as
\begin{equation}
\abovedisplayshortskip=2pt
\belowdisplayshortskip=2pt
\abovedisplayskip=2pt
\belowdisplayskip=2pt
\begin{aligned}
\bm{X}_{i,\mathrm{cft}}\left(f\right)&=\bm{X}_{i,\mathrm{dtft}}\left(\frac{fT}{Q}\right)=\bm{X}_{i,\mathrm{dtft}}\left(\frac{f}{Q{f}_B}\right)
\\
&=\sum_{l=1}^{L}{\bm{f}_lS_u\left(\frac{f}{Q{f}_B}\right)}D_{Q\zeta_l}\left(\frac{f}{Q{f}_B}\right)
\\
&=\sum_{l=1}^{L}{\bm{f}_lS\left(Q\frac{f}{Q{f}_B}\right)}D_{Q\zeta_l}\left(\frac{f}{Q{f}_B}\right)
\\
&\overset{(a)}{\approx}\sum_{l=1}^{L}{\bm{f}_lS\left(\frac{f}{f_B}\right)e^{-j2\pi fT\zeta_l}},
\\
&f\in\left[-\frac{Qf_B}{2}\right.,\left.\frac{Qf_B}{2}\right),
\end{aligned}
\label{cft_xi}
\end{equation}
where $S_u(f)$ and $S(f)$ are the DTFT of $s_u[n]$ and $s[n]$ respectively, and (a) holds due to (\ref{frac_filter}). Then, the CFT of the transmitted signal is
\begin{equation}
\abovedisplayshortskip=2pt
\belowdisplayshortskip=2pt
\abovedisplayskip=2pt
\belowdisplayskip=2pt
\begin{aligned}
\bm{X}\left(f\right)&=\bm{X}_{i,\mathrm{cft}}\left(f\right)\mathrm{\varPhi}\left(f\right)
\\
&\approx\sum_{l=1}^{L}{\bm{f}_lS\left(\frac{f}{f_B}\right)\mathrm{\varPhi}\left(f\right)e^{-j2\pi fT\zeta_l}},
\\
&f\in\left[-\frac{Qf_B}{2}\right.,\left.\frac{Qf_B}{2}\right),
\end{aligned}
\label{cft_x}
\end{equation}
where $\mathrm{\varPhi}\left(f\right)$ is the CFT of $\varphi\left(t\right)$. It is observed from (\ref{cft_x}) that the bandwidth of the transmitted signal $\bm{X}\left(f\right)$ is limited by that of $\mathrm{\varPhi}\left(f\right)$. As a result, even though upsampling is performed at the transmitter, the bandwidth of the transmitted signal does not increase, which is desired. Note that for $f\in\left[-0.4Qf_B\right.,\left.0.4Qf_B\right)$, the approximation in (\ref{cft_x}) is accurate. The upsampling factor $Q$ can then be chosen according to the bandwidth of $\mathrm{\varPhi}\left(f\right)$. Let $B_{\mathrm{\varPhi}}$ denote the double-sided bandwidth of $\mathrm{\varPhi}\left(f\right)$, then $Q$ should be chosen so that
\begin{equation}
\abovedisplayshortskip=2pt
\belowdisplayshortskip=2pt
\abovedisplayskip=2pt
\belowdisplayskip=2pt
\begin{aligned}
0.8Qf_B \geqslant B_{\Phi}\,\text{ or }\,Q\geqslant 1.25\frac{B_{\Phi}}{f_B}.
\end{aligned}
\label{choice of Q}
\end{equation}
Typically, for the raised cosine pulse with roll-off factor $\beta$, we have $B_{\Phi}=(1+\beta)f_B$. Thus,  (\ref{choice of Q}) can be further written as
\begin{equation}
\abovedisplayshortskip=2pt
\belowdisplayshortskip=2pt
\abovedisplayskip=2pt
\belowdisplayskip=2pt
\begin{aligned}
Q\geqslant 1.25\left( 1+\beta \right).
\end{aligned}
\label{choice of Q raised cosine}
\end{equation}
In particular, for $\beta\leq0.6$, an upsampling factor $Q=2$ would suffice. From (\ref{cft_x}), by using inverse continuous Fourier transform (ICFT), the time-domain transmitted signal of the proposed fDAM is given by
\begin{equation}
\abovedisplayshortskip=2pt
\belowdisplayshortskip=2pt
\abovedisplayskip=2pt
\belowdisplayskip=2pt
\begin{aligned}
\bm{x}\left(t\right)&\approx\sum_{n=-\infty}^{\infty}\sum_{l=1}^{L}{\bm{f}_ls\left[n\right]\varphi\left(t-\zeta_lT-nT\right)}
\\
&=\sum_{l=1}^{L}{\bm{f}_l\sum_{n=-\infty}^{\infty}s\left[n\right]\varphi\left(t-nT-\zeta_lT\right)}
\\
&=\sum_{l=1}^{L}{\bm{f}_ls\left(t-\zeta_lT\right)},
\end{aligned}
\label{x_t_frac}
\end{equation}
where $s\left(t\right)$ is the information-bearing symbols after pulse shaping as defined in (\ref{s_t}). 
Thus, with the channel impulse response given in (\ref{h_t}), the received signal is
\begin{equation}
\abovedisplayshortskip=2pt
\belowdisplayshortskip=2pt
\abovedisplayskip=2pt
\belowdisplayskip=2pt
\begin{aligned}
y\left(t\right)&=\bm{h}^H\left(t\right)\ast\bm{x}\left(t\right)+z\left(t\right)
\\
&\approx\sum_{l=1}^{L}{\bm{h}_l^H\bm{f}_ls\left(t-\zeta_lT-\tau_l\right)}
\\
&+\sum_{l=1}^{L}\sum_{l^\prime\neq l}^{L}{\bm{h}_l^H\bm{f}_{l^\prime} s\left(t-\zeta_{l^\prime}T-\tau_l\right)}+z\left(t\right).
\end{aligned}
\label{y_t_frac}
\end{equation}

By letting $\zeta_l=\frac{\tau _{\max}-\tau _l}{T}$, we have
\begin{equation}
\abovedisplayshortskip=2pt
\belowdisplayshortskip=2pt
\abovedisplayskip=2pt
\belowdisplayskip=2pt
\begin{aligned}
y\left(t\right)&\approx\left(\sum_{l=1}^{L}{\bm{h}_l^H\bm{f}_l}\right)s\left(t-\tau_{\max}\right)
\\
&+\sum_{l=1}^{L}\sum_{l^\prime\neq l}^{L}{\bm{h}_l^H\bm{f}_{l^\prime}s\left(t-\tau_{\max}+\tau_{l^\prime}-\tau_l\right)}+z\left(t\right).
\end{aligned}
\label{y_t_frac_2}
\end{equation}
Thus, with the path-based ZF beamforming, the received signal can be expressed as
\begin{equation}
\abovedisplayshortskip=2pt
\belowdisplayshortskip=2pt
\abovedisplayskip=2pt
\belowdisplayskip=2pt
\begin{aligned}
y\left(t\right)\approx\left(\sum_{l=1}^{L}{\bm{h}_l^H\bm{f}_l}\right)s\left(t-\tau_{\max}\right)+z\left(t\right).
\end{aligned}
\label{y_t_frac_3}
\end{equation}

It is observed from (\ref{y_t_frac_3}) that similar to the original iDAM design under the assumption of integer multi-path delays, the proposed fDAM can transform the multi-path channel with fractional delays to a simple AWGN channel with a single delay $\tau_{\max}$. Besides, all the $L$ multi-path signal components will contribute to the desired signal power. 

\section{Simulation Results}\label{Sim}
Simulation results are provided to evaluate the performance of the proposed fDAM. The number of transmit antennas is $M = 64$, the carrier frequency is $f_c = 28 \,\mathrm{GHz}$, and the symbol interval is $T = 2.5\,\mathrm{ns}$. Note that since no ISI exists at the receiver for the proposed fDAM, sinc pulses can be used for maximum spectral efficiency. However, since sinc pulse filters cannot be implemented precisely, a raised cosine pulse filter with a roll-off factor of $\beta=0.05$ is used in the simulation. The bandwidth of the signal is thus $B=(1+\beta)f_B=420\,\mathrm{MHz}$, where $f_B=\frac{1}{T}=400\,\mathrm{MHz}$ is the symbol rate. The number of multi-paths is $L = 3$. For the scenarios with fractional multi-path delays, the delays are chosen randomly from the continuous interval $[0, 200T]$. On the other hand, for the simplified scenarios by assuming integer multi-path delays, the above randomly generated delays are rounded so that all the delays are integer multiples of $T$. 

For the benchmarking OFDM scheme, the number of subcarriers is $N_{sc}=1024$, and the CP length is $N_{cp}T=200T$, which is no smaller than the maximum possible delay spread. The channel coherence time is  $T_c = 1 \,\mathrm{ms}$. For fairness, the time-domain signal after OFDM modulation is passed through the same transmit pulse shaping filter used in DAM. This ensures that both systems transmit with the same bandwidth limit. 16-QAM constellation is used for both systems. The received SNR varies from $0 \mathrm{dB}$ to $20 \mathrm{dB}$. The following results are averaged over 500 channel realizations and the symbol error rates (SERs) are calculated over 1024 symbols per channel realization.

\begin{figure}[!htbp]
\vspace*{-10pt}
\centering
\setlength{\abovecaptionskip}{0pt}
\setlength{\belowcaptionskip}{0pt}
    \includegraphics[width=0.4\textwidth]{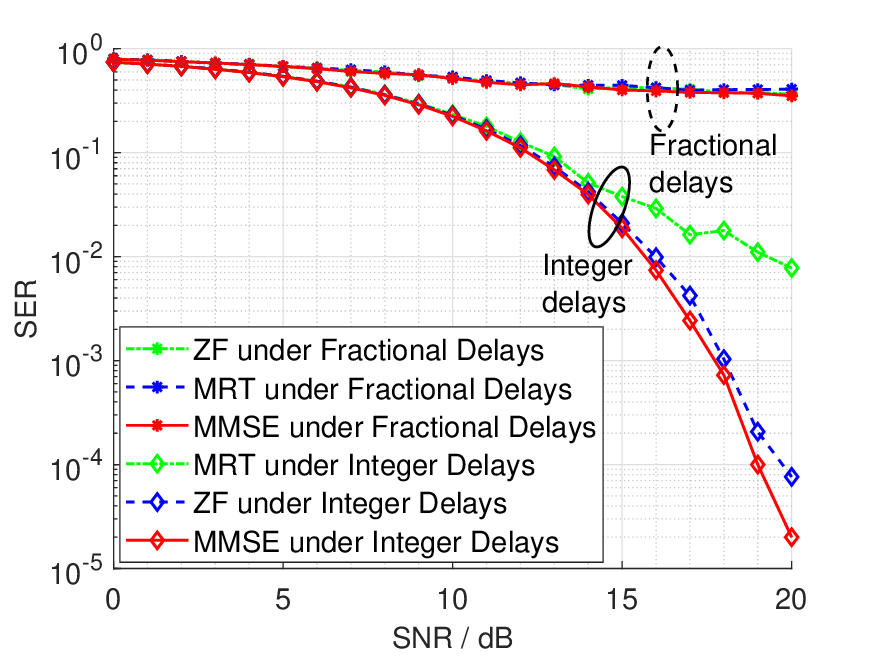}
	\hfil
\caption{SER of the existing iDAM design for scenarios with integer and fractional multi-path delays.}
	\label{fig_effect_frac}
 \vspace*{-10pt}
\end{figure}

Firstly, the impact of fractional delays on the existing iDAM is evaluated. Fig. \ref{fig_effect_frac} shows the resulting SER of the iDAM design in \cite{lu2022delay}\cite{lu2023delay}. It is observed that, although iDAM gives rather good performance under integer multi-path delays, the existence of fractional delays can cause severe performance degradation, regardless of the beamforming schemes used. In particular, due to the residual ISI caused by fractional delays, iDAM results in SER greater than $10^{-1}$. Since the performance of ZF and MMSE are similar, in the following, ZF beamforming is used for further comparison between iDAM and the proposed fDAM.

\begin{figure}[!htbp]
\vspace*{-10pt}
\centering
\setlength{\abovecaptionskip}{0pt}
\setlength{\belowcaptionskip}{-3pt}
    \includegraphics[width=0.4\textwidth]{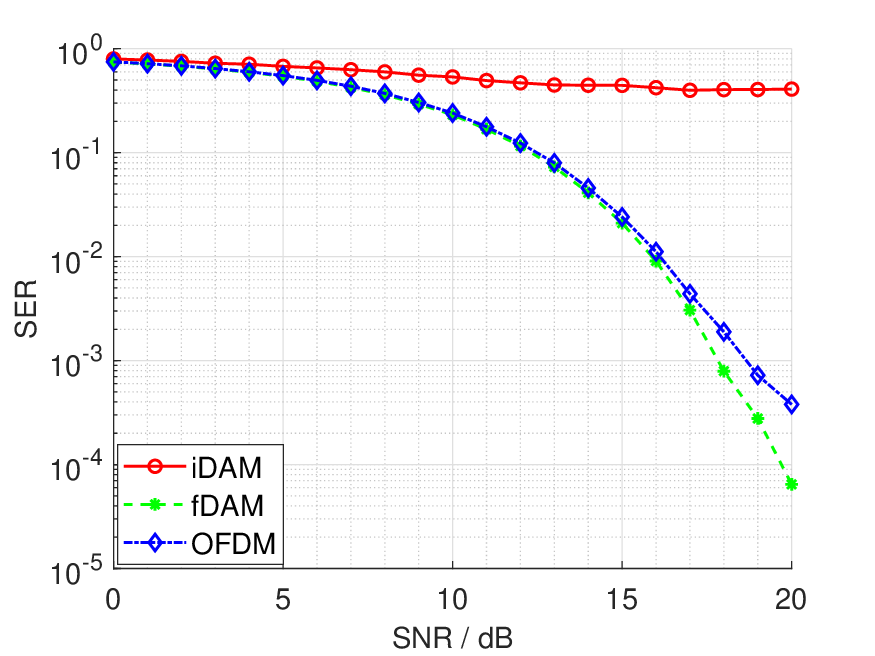}
\caption{SER of iDAM, fDAM and OFDM for channels with fractional multi-path delays.}
	\label{fig_fDAM_OTDAM_ser}
 \vspace{-6pt}
\end{figure}

Fig. \ref{fig_fDAM_OTDAM_ser} compares the SER of the proposed fDAM to the conventional iDAM and OFDM for channels with fractional multi-path delays. It is observed that under the more general and practical scenarios with fractional multi-path delays, fDAM can significantly outperform iDAM, thanks to its effective elimination of residual ISI brought by the fractional delays. Note that for OFDM, the equivalent channel after the subcarrier-based maximal-ratio transmission (MRT) beamforming can still be frequency selective even with $M=64$ antennas, causing different SNRs among subcarriers. By contrast, the fDAM results in constant SNR that is contributed by all the $L$ multi-path components, as can be seen in (\ref{y_t_frac_3}).  Additionally, the transmit and receive filters of OFDM can cause attenuation for subcarriers at the edge of the frequency band. As a result, the proposed fDAM may even slightly outperform OFDM. Note that the SER overhead caused by CP \cite{lu2023delay} is not considered here, with which the advantage would be greater.

\begin{figure}[!htbp]
\vspace*{-12pt}
\centering
\setlength{\abovecaptionskip}{0pt}
\setlength{\belowcaptionskip}{0pt}
    \includegraphics[width=0.4\textwidth]{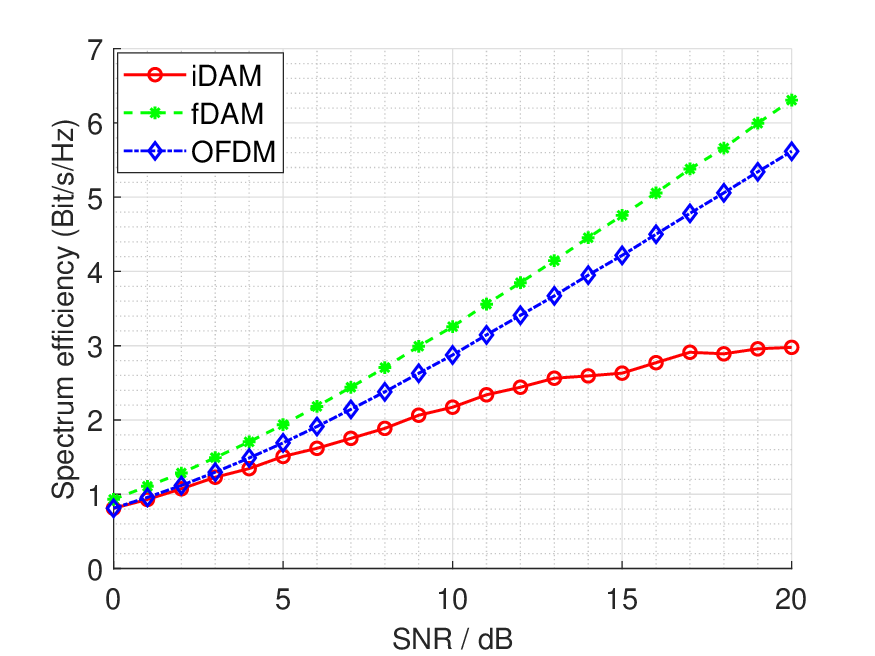}
\caption{Spectrum efficiency of iDAM, fDAM and OFDM for channels with fractional multi-path delays.}
	\label{fig_fDAM_OTDAM_achievable_rate}
 \vspace*{-10pt}
\end{figure}

The spectral efficiency of iDAM, fDAM and OFDM is shown in Fig. \ref{fig_fDAM_OTDAM_achievable_rate}. Under fractional multi-path delays, since fDAM can greatly improve SINR by completely eliminating the ISI, the achievable rate of fDAM is considerably larger than that of iDAM. Besides, the proposed fDAM also achieves significantly larger spectrum efficiency than OFDM. This is mainly due to the saving of guard interval overhead. Specifically, for fDAM, the overhead introduced by the guard interval between different channel coherence blocks is very small \cite{zhang2023delay}, and the overhead here is mainly determined by the roll-off factor of the pulse shaping filter, which is around $\frac{\beta}{(1+\beta)} = 4.8\%$. This can be further lowered by reducing the roll-off factor, and the limit is close to 0. For OFDM, the main overhead is caused by CP, which can be calculated as $\frac{N_{cp}}{N_{sc}+N_{cp}}=16.3\%$. Note that the overhead for OFDM can theoretically be lowered to near 0 by increasing $N_{sc}$ so that $(N_{sc}+N_{cp})T = T_c $, but this would mean $N_{sc} = 399800$ subcarriers, which is impractical since it would lead to extremely high PAPR and vulnerability to frequency offsets.

\begin{figure}[!htbp]
\vspace{-10pt}
\centering
\setlength{\abovecaptionskip}{0pt}
\setlength{\belowcaptionskip}{0pt}
    \includegraphics[width=0.4\textwidth]{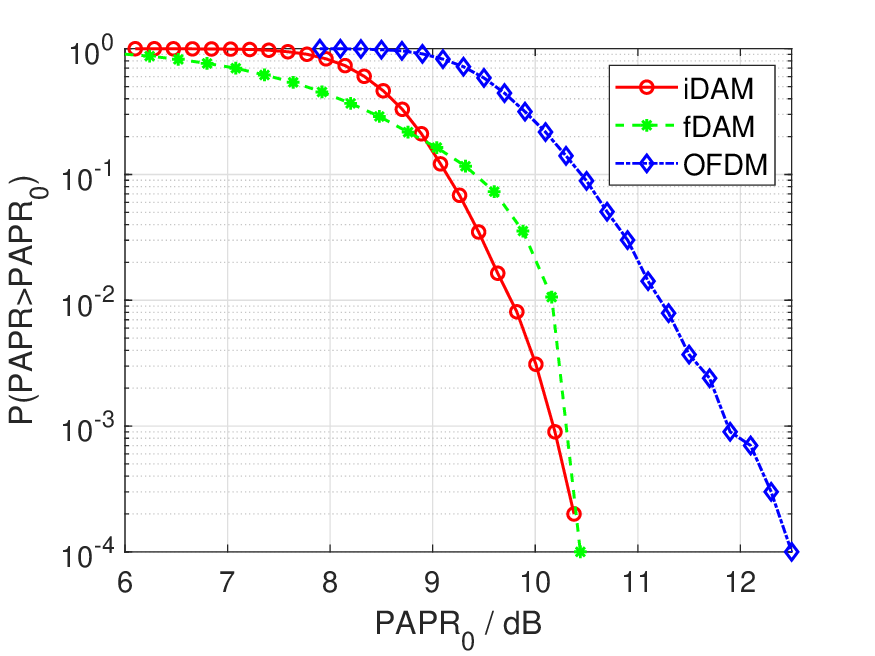}
\caption{PAPR of iDAM, fDAM and OFDM.}
	\label{fig_fDAM_OTDAM_PAPR}
 \vspace*{-12pt}
\end{figure}

The PAPR of iDAM, fDAM and OFDM is compared in Fig. \ref{fig_fDAM_OTDAM_PAPR}. Different from the PAPR calculated in \cite{zhang2023delay}, here, the PAPR is calculated in the analog domain after transmit filtering. As shown in Fig. \ref{fig_fDAM_OTDAM_PAPR}, the PAPR of fDAM is slightly worse than that of iDAM, but their upper limit is approximately the same. OFDM has the worst PAPR among the three, which is expected since it is a multi-carrier transmission technique. Note that the gap between OFDM and iDAM is smaller compared to previous work \cite{zhang2023delay}. This is due to the aforementioned transmit filtering, without which the gap would be significantly larger.

\section{Conclusion}
This paper first analyzed the impact of fractional multi-path delays on existing iDAM techniques, and then a novel method called fDAM was proposed to achieve perfect delay alignment by leveraging upsampling with factor $2$ and fractional delay filters.  Simulation results demonstrated the superior performance of the proposed fDAM technique over the existing iDAM design and the conventional OFDM in terms of SER, spectral efficiency and PAPR.

\bibliographystyle{IEEEtran}
\bibliography{IEEEabrv,reference.bib}

\end{document}